\documentclass[useAMS,usegraphicx,usenatbib]{mn2e}
\newcommand{\msun}{M$_\odot$}
\title[Constraining  the mass of the GRB 030329 progenitor]{Constraining  the mass of the GRB 030329 progenitor\thanks{Based on observations with the NASA/ESA {\it Hubble Space 
Telescope}, obtained at the Space Telescope Science Institute, which is 
operated by the association of Universities for Research in 
Astronomy, Inc., under NASA contract NAS5-26555.}}
\author[G. \"Ostlin et al.]{G\"oran \"Ostlin$^{1}$, Erik Zackrisson$^{2,1,3}$, Jesper Sollerman$^{1,4}$, Seppo Mattila$^{5,2}$,  Matthew Hayes$^{1,6}$ \\
$^1$Stockholm Observatory, Department of Astronomy, Stockholm University, AlbaNova University Center,  106 91, Stockholm, Sweden\\
$^2$Tuorla Observatory, University of Turku, V\"ais\"al\"antie 20, FIN-21500 Piikki\"o, Finland\\
$^3$Department of Astronomy and Space Physics, Box 515, SE-75120 Uppsala, Sweden\\
$^4$Dark Cosmology Centre, Niels Bohr Institute, University of Copenhagen, Juliane Maries Vej 30, 2100 Copenhagen, Denmark\\ 
$^5$Astrophysics Research Centre, School of Mathematics and Physics, Queen's University Belfast, Belfast BT7 1NN, UK \\
$^6$Geneva Observatory, University of Geneva, 51 chemin des Maillettes, 1290 Sauverny, Switzerland}
\begin{document}
\date{Accepted ... Received ...; in original form ...}
\pagerange{\pageref{firstpage}--\pageref{lastpage}} \pubyear{2007}
\maketitle
\label{firstpage}

\begin{abstract}
The long-duration gamma-ray burst (GRB) 030329, associated with supernova (SN) 2003dh,
occurred inside a star-forming dwarf galaxy at redshift $z=0.1685$. The low redshift, and a 
rich set of archival Hubble Space Telescope (HST) images, makes this GRB  well-suited for a detailed study
of the stellar population in the immediate vicinity of the explosion. Since the lifetime of 
a star is directly tied to its mass, the age of the stellar population can be used to put
constraints on the GRB and SN progenitor mass. 
From the HST images  we 
extract the colours of the precise site from which the GRB originated, and find that the 
colours  are significantly different from those of the overall host galaxy
and the surrounding starburst environment.

We have used spectral evolutionary models, including nebular emission, 
to carefully constrain the age of the stellar population, and hence the 
progenitor, at the very explosion site. 
 For instantaneous burst models we find that a population age of $\sim$5 Myr  
 best matches the data, suggesting a very massive ($M_\mathrm{ZAMS} > 50 
 \ M_\odot$) star as the progenitor, with an upper limit of 8 Myr 
 ($M_\mathrm{ZAMS} > 25 \ M_\odot$).
For more extended star formation scenarios, the inferred progenitor age is 
in most cases still very young ($< 7$  Myr, implying $M_\mathrm{ZAMS} > 25 
\ M_\odot$), with an upper limit of 20 Myr ($M_\mathrm{ZAMS} > 12 \ M_\odot$).

These age estimates are an order of magnitude lower than the 
ages inferred from the overall host galaxy colours, indicating that progenitor mass estimates 
based on 
data for spatially unresolved GRB host galaxies will in 
general be of limited use. Our results are consistent with the collapsar scenario.
\end{abstract}
\maketitle

\section{Introduction}

It is now well established that long duration GRBs are connected to the deaths of massive stars. 
The standard picture is reviewed by \citet{woosleybloom06}, where the prevailing model is a 
collapsar \citep{macfadyenw}, i.e. a massive star collapsing to a black hole and producing an 
energetic supernova. 
The  first piece of evidence for the supernova scenario came with the discovery of SN\,1998bw 
in the error box of GRB\,980425 \citep{galama98,patat01}, and was later confirmed by the association
of GRB\,030329  with SN\,2003dh \citep{hjorth03,stanek03,matheson03}.
It is generally 
assumed that the progenitors to these GRBs  are very massive stars. 
Not all types of core-collapse SNe produce 
GRBs, and it is thought that only the upper mass range of the IMF is contributing to the GRBs. 
Moreover, the modeling of the properties of the  supernovae accompanying the long 
GRBs points to very massive progenitors  \citep[e.g.][]{woosleybloom06}.
In the case of SN\,2003dh, which was associated with GRB\,030329, estimates in the range of
25 to 40 $M_\odot$ has been obtained from modelling  the supernova observations \citep{mazzali03,deng05}.

Arguments for a massive stellar progenitor for long-duration GRBs also come from studies 
of the host galaxies.
 \citet{fruchter06} confirmed that the GRBs occur in blue star-forming galaxies, and 
are concentrated to the most active star forming regions. For some nearby supernovae not 
connected to GRBs, a study of the actual progenitor star has been possible. Recent 
programmes have taken a  systematic approach to this question and have revealed several 
supernova progenitor stars in pre-explosion images \citep[e.g.][]{2004Sci...303..499S,2005PASP..117..121L,2007ApJ...656..372}. However, all the direct detections so far
are for type II SNe with moderately massive progenitor stars. Furthermore,
the progenitors of type Ib/c SNe still remain to be directly detected
\citep[e.g.][]{2005ApJ...630L..33M,2007MNRAS.381..835C}
with the exception of SN 2006jc 
for which an 
outburst was observed 
coincident with the SN just two years before its explosion 
\citep{pastorello,2007ApJ...657L.105F}. 
For GRBs, which are typically more distant, we cannot detect single stars prior to explosion. An alternative way to probe 
the progenitor is to analyse the population of stars near the explosion site. Such modeling 
of the entire host galaxies has revealed that, overall,  the stellar populations of GRB hosts are dominated by 
young stars. However, to put constraints on the actual GRB progenitor requires a close look at the 
very site where the GRB was born and exploded.  A population study of 
the environment of GRB\,980425 was carried out by \citet{sollerman05}, who concluded that this GRB 
originated in a star more massive than 30 $M_\odot$. More recently, \citet{thone}  found 
support for a massive star origin of the GRB\,060505 which, however, did not have an associated
luminous supernova.
These results are consistent with the standard 
picture, although further and tighter constraints should be sought.

To probe the immediate environment of the explosions requires nearby bursts and/or high-resolution 
imaging. In this paper we investigate archival Hubble Space Telescope (HST) images obtained of the 
host of GRB\,030329 at $z=0.1685$ \citep{gorosabel05}.  The host   appears to be a subluminous 
($L\sim 0.016L^\star$), metal-poor, galaxy with a star formation rate of $\sim 0.5 \ M_\odot$ yr$^{-1}$ 
and a total stellar mass of $M\sim 10^8$--$10^9 \ M_\odot$, dominated by a rather young stellar 
population \citep{gorosabel05,sollerman05}. The host galaxy will be further discussed in Sect. 4.2. 

In this paper, we use the HST images to zoom in on the exact location of the GRB explosion. 
We show that useful constraints on the progenitor mass can be obtained by careful modeling 
of the colours as seen by HST. Such studies are complementary to e.g. modeling of the supernova 
component -- and will help to constrain the properties of the exploding star.

In Sect.~\ref{observations} we describe the observational material
 used. In Sect.~\ref{analysis} we describe how the optical data were analysed, i.e. how we 
subtracted the old population from the young star burst region. In Sect.~\ref{model} we describe 
the modeling of the stellar populations  and discuss the constraints we derive on the progenitor. 
In Sect. \ref{outlook} we discuss how
age estimates can be improved and in Sect. \ref{conclusion} we summarise our conclusions.

Throughout the paper we are assuming the following cosmology:
$\Omega_\mathrm{M}=0.3$, $\Omega_\Lambda=0.7$, $H_0=72$ km s$^{-1}$ Mpc$^{-1}$.


\section{Observations}
\label{observations}

All the data analysed in this paper have been retrieved from the Hubble Space 
Telescope (HST) data archive. Some of the images have previously been discussed by 
\citet{sollerman05} and \citet{fruchter06}.


\subsection{The ACS/WFC data}

GRB\,030329 was observed with the Wide Field Camera (WFC) of the Advanced Camera for Surveys (ACS) 
on several occasions between about two weeks after the
explosion and more than two years subsequently. Since we are  here interested
in the properties of the host galaxy, early  images 
are of little value except for pinpointing the exact GRB location. The images used in
this paper are summarised in Table \ref{data}. 

\begin{figure*}
\includegraphics[width=184mm]{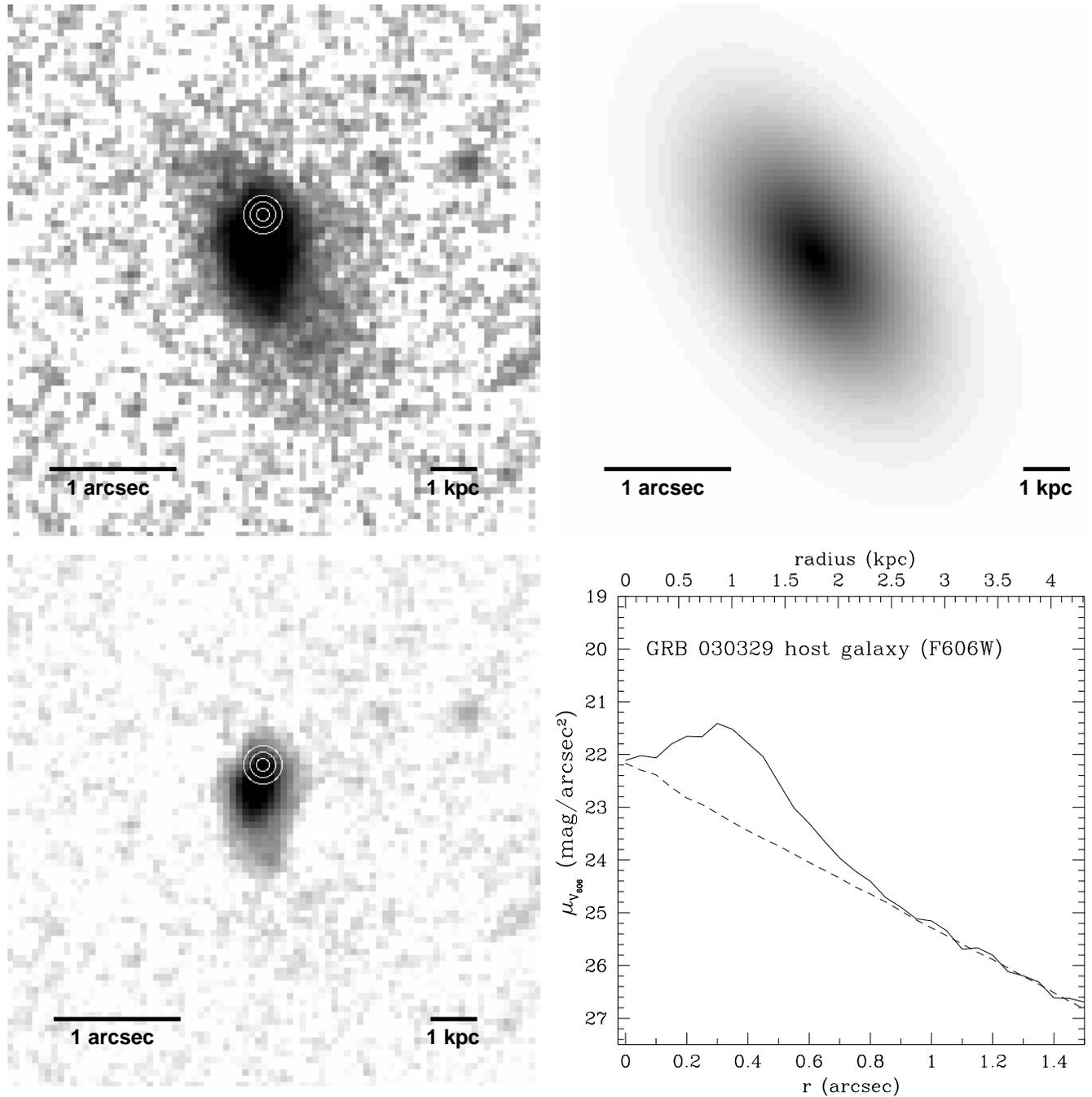}
\caption[]{
{\bf Upper left:}
The host galaxy of GRB 030329 in the F606W filter (HST/ACS). The field shown is
$4\farcs2 \times 4\farcs2$  which corresponds to $12\times 12$ kpc at the
distance of GRB\,030329. This image is made from data obtained at different
epochs and is not contaminated by light from the GRB afterglow. The location
of the GRB explosion is indicated by the concentric white circles (where the
innermost circle corresponds to the used photometric aperture). 
{\bf Upper right:}
Synthetic exponential disk model fitted to the outer isophotes of the deep
F606W image. The inclination is $55^\circ$ and the scale length $0\farcs31$
(0.87 kpc).
{\bf Lower left:}
Here we show the result of subtracting the disk model from the F606W image in the
upper left panel. This has  isolated a central starburst region, free of
contamination from the underlying population described by the best fit disk model.
The GRB occurred near the edge of the central starburst region,
$\sim 0\farcs2$ from its centre.
{\bf Lower right:} The luminosity profile resulting from integrating the 
unsubtracted F606W
image in the upper left panel in concentric ellipses (with centroid from the outer 
isophotes) is shown as a solid line
while the fitted disk model is shown as a dashed line.
}
\label{composite}
\end{figure*}

The drizzled F606W images were aligned to a common origin
We co-added these images sorted  in four temporal bins: 
'early', 'mid', 'late' , and 'very late' (see Table \ref{data}).
The 'early' images are useful for identifying the precise location of the GRB, 
but the afterglow contaminates the light from the host galaxy. In the 'mid'
images, the afterglow has faded enough that these data can be used to study 
the host galaxy far from the explosion site. The 'very late' images contain no 
detectable afterglow  or supernova light, 
and from comparing 'late' and 'very late' images, we 
conclude that 'late' images are safe to use also for the GRB environment.
We made a total  GRB-free image by masking out the GRB affected 
regions in the mid images and adding these to the late and very late images.
In this way the final image has  maximum depth in the outskirts of the host
galaxy, but no afterglow  or supernova contamination. 
The resulting image is shown in 
the upper left panel of Fig.~\ref{composite}.
We used the same strategy to construct a GRB free image in F814W. For F435W 
we used the co-added images from a single 'late' dataset. 

Photometric zero points were taken from the ACS www-pages\footnote{http://www.stsci.edu/hst/acs} 
and are: 25.779,  26.398, 25.501  for the F435W, F606W and F814W filters, respectively, 
all in the {\sc vegamag} system, which will be adopted throughout this paper. 
In what follows we shall refer to magnitudes in this system as $B_{435}$, $V_{606}$ and $I_{814}$.
All data that is presented has been corrected for a foreground Galactic 
reddening of $E(B-V) = 0.025$~mag \citep[][]{schlegel}. 
Taking the different bandpass of the used filters with respect to the
'standard' system into account, the applied  corrections 
are (in magnitudes): 
$A(B_{435})=0.109, ~A(V_{606})=0.075, ~A(I_{814})=0.049$.

\begin{table}
   \begin{center}
      \caption[]{ACS/WFC images used (exposure time in seconds).
}
      \label{data}
      \begin{tabular}{lllrc}
\hline
Filter & Date & Dataset name & Exp. time & Class \\
\hline
F606W & Apr. 15, 2003  & J8IY2B021  & 400 &  early \\  
F606W & Apr. 21, 2003  & J8IY2J021  & 400 &  early \\ 
F606W & May 12, 2003  & J8IY2M021  & 450 &  early \\ 
F606W & Nov. 12, 2003  & J8IY2T010  & 1920 &  mid \\ 
F606W & Nov. 12, 2003  & J8IY2T030 & 2088 &  mid  \\ 
F606W & Feb. 4, 2004  & J8IY2U010 & 1920 &  mid  \\ 
F606W & May 25, 2004 & J8IY9Z010 & 1920 & late \\ 
F606W & May 25, 2004 & J8IY9Z020 & 2080 & late \\ 
F606W & May 9, 2005 & J8IYF2010 & 1920 & very late \\ 
F606W & May 9, 2005 & J8IYF2030 & 2080 & very late \\ 
\hline
F435W & May 24, 2004 & J8IY9V010 & 1920 & late \\
\hline
F814W & Nov. 12, 2003 & J8IY2T050 & 2040 & mid \\
F814W & May 24, 2004 & J8IY9V020 & 2040 &  late  \\
\hline
      \end{tabular}
   \end{center}
\end{table}

\subsection{The ACS/HRC/F250W data}
The archive also contains several observations in the F250W filter with the 
UV-sensitive High Resolution Camera (HRC) of ACS, which were 
also retreived. However,
we could not properly align the HRC/F250W observation with those obtained
with the WFC. The different apertures of the two cameras results in slight
inconsistencies in the astrometric keywords.
As a result, we could not be sure that the {\sc Multidrizzle} software had
correctly registered the images at the requested pixel-level accuracy. 
Remaining misalignments could not be fixed by hand  
 since
no for- or background point-sources 
were available for the image registration. Hence, we did not
use the F250W data in our final analysis.

\subsection{NICMOS data}
GRB\,030329 was also observed  with the NIC2 camera of the NICMOS instrument onboard HST.
These data suffered from well known artefacts in the form of non-linear temporal bias 
drifts during the 
exposures\footnote{see http://www.stsci.edu/hst/nicmos}. 
The data were re-reduced using the CALNICA pipeline and the
artifacts removed using the BIASEQ and PEDSKY
tasks of STSDAS. Unfortunately, the latest epochs which 
were the only ones free of GRB contamination suffered from cosmic ray 
persistence limiting the depth of these observations. Aligning the  NIC2
images with ACS/WFC  was complicated due to the 
very few 
reference sources  visible within the  NIC2 field of view.
We therefore rotated the images to the  orientation of the WFC data 
and used the GRB itself, 
present in the early NIC2 images, for centring. The images potentially
 relevant for our analysis are made up of two exposures in each of the
 F110W and F160W filters, taken on May 23, 2004. The total 
exposure time in each filter is 1216 s. However, only the F110W 
image turns out to be useful for our purposes, as further discussed in 
Sect.~\ref{analysis}. The adopted NIC2+F110W zeropoint is 22.95 
({\sc Vegamag}) and magnitudes  will hereafter be denoted 
by $J_{110}$.

\subsection{HST spectroscopic data}

GRB\,030329 was observed with both STIS and ACS in spectroscopic mode. 
It could potentially be very useful to constrain the internal extinction
from the H$\alpha$/H$\beta$ ratio at the actual location of the  burst. 
However, these 
spectra\footnote{http://www-int.stsci.edu/$\sim$fruchter/GRB/030329/index.html} 
do not have enough signal, or are too heavily
 contaminated by the GRB afterglow at early epochs, to allow such an analysis.

\section{Photometric analysis}
\label{analysis}

We examined the shape of the host galaxy in the final very deep F606W image. We fitted ellipses
to the outer isophotes and derived a luminosity profile by integrating in elliptic annuli.
The structure of the outskirts of the host galaxy is well fitted
by an exponential disk with scale length of $0\farcs3$ (0.9 kpc) and an inclination angle of $i=55\degr$ (assuming $\cos(i)=b/a$, i.e. an infinitely thin disk). 
   We created a synthetic galaxy image with these properties, scaled to the 
surface brightness level of the outer parts of the host galaxy and  subtracted it from the host
galaxy image. In this way we obtained an image of the central starburst with no 
contamination from the underlying galaxy. The same method was used to obtain images
of the central starburst in the F435W and F814W filters. Being cleaned from the underlying disk
contribution, these images are suitable for 
a photometric modeling of the properties of the GRB explosion environment. 
In Fig.~1 we show the
luminosity  profile of the host and the resulting image after subtracting 
the underlying host galaxy.

We then proceeded to extract colours for the slightly off-centre 
GRB explosion site. In order to check how sensitive
the results are to the host galaxy subtraction, we increased the luminosity of the disk
model to the upper limit ('max disk') allowed by the uncertainties, and also derived the 
colours  without subtracting an underlying disk  ('no disk').

This photometric method is appropriate if the stellar population from which the 
GRB originated dominates the light within the aperture. 
An alternative method is to also subtract the median flux in a surrounding annulus, 
as  is done in conventional aperture photometry, and this method was
used as a consistency check.

The GRB environment photometry is presented in Table 2, and was extracted for an 
aperture radius of $r=0\farcs1$.
We also experimented with a smaller aperture which gave consistent results. 
For the $r=0\farcs1$ photometry, aperture corrections of 0.510, 0.461 and 
0.538 mag were subtracted 
from $B_{435}$, $V_{606}$ and $I_{814}$, respectively \citep{sirianni}.

The scaling of the underlying disk has a minor impact
on the colours of the GRB explosion site. The statistical  uncertainties 
are estimated to be $\sigma_B=0.03, \sigma_V=0.015, \sigma_I=0.035$ magnitudes. The 
differences in Table 2 give a good idea about the systematic uncertainties, in which 
the subtraction (or not) of the local background has the largest impact. 
By modeling the stellar population (see Sect. \ref{discussion}) using the 
full range of values in Table 2, we also include the systematic uncertainties 
 in our analysis.

In addition, we present in Table 3, the photometry  for the whole galaxy.

\begin{table*}
   \begin{center}
      \caption[]{GRB environment photometry. The upper entries are for
photometry extracted in a $0\farcs1$ radius aperture without subtracting
the local background. For the lower entries, local background has been 
subtracted using the median value in one pixel wide annulus immediately
outside the photometric aperture. All magnitudes are in the HST VEGAMAG 
system.
}
      \label{tab-regions}
      \begin{tabular}{llrrrrr}
\hline
Disk  & Local& $B_{435}$ & $V_{606}$ & $I_{814}$ & $B_{435}-V_{606}$ & $V_{606}-I_{814}$ \\
model & backgr. \\
\hline
%
%
Max disk     & no & 24.843 & 24.778 &  24.237   & 0.066 & 0.540 \\ 
Best disk & no & 24.816 & 24.766 &  24.204   & 0.050 & 0.561 \\ 
No disk      & no & 24.729 & 24.679 &  24.082   & 0.051 & 0.596 \\ 
\hline
Max disk     & yes & 25.126 & 25.156 & 24.696 & --0.044 & 0.480 \\ 
Best disk & yes & 25.114 & 25.159 & 24.678  & --0.029 & 0.460 \\ 
No disk      & yes & 25.117 & 25.116 & 24.610 & 0.002 & 0.506 \\ 
\hline
      \end{tabular}
   \end{center}
\end{table*}

%

 \subsection{NICMOS data}
We were not able to clearly detect and fit the underlying 
host galaxy in the NICMOS images, but 
have extracted the flux in the GRB-region in apertures matching those of the WFC data. 
 In the F160W 
image  many pixels near the GRB explosion site are negative and we regard these
data as unreliable and will not use them further. In the F110W  image we could 
extract a magnitude for the same region as used for the WFC data
but without subtraction of the underlying galaxy (corresponding to the 'no disk'
option). In this way we obtained  $J_{110}=24.9 \pm 0.4$ mag., 
where we have adopted the definition of total NICMOS magnitudes and  an aperture 
correction derived from the encircled energy 
distribution given in the HST/NICMOS Instrument Handbook \citep{nicmosihb}.

\begin{table}
   \begin{center}
      \caption[]{Integrated host galaxy photometry within the Holmberg radius (1.4\arcsec, equivalent
to 3.9 kpc) 
}
      \label{tab-regions}
      \begin{tabular}{ccccc}
\hline
 $B_{435}$ & $V_{606}$ & $I_{814}$ & $B_{435}-V_{606}$ & $V_{606}-I_{435}$ \\
\hline
23.2    & 22.8  & 22.3 & 0.36 & 0.52  \\ 
\hline
      \end{tabular}
   \end{center}
\end{table}

\section{Stellar Population Analysis}\label{model}\label{modeling}

We will now describe how we use spectral evolutionary synthesis
models together with the derived photometry to estimate the age of the stellar
population. Once an age estimate has been found, this can be converted into a lower limit on the 
zero age main sequence progenitor mass ($M_\mathrm{ZAMS}$) using 
 stellar evolutionary models from  \citet{meynet} and 
\citet{fagotto}. The lifetimes for these two independent models agree well,
typically to within 5\%. More recent calculations \citep{hirschi} that include 
stellar rotation indicate that,  at least for solar metallicity,  
the effects on the life times are insignificant. 
 For hypothetical, extremely rapidly rotating stars \citep[e.g.][]{woosleyheger06}, 
the mass lifetime--relation could possibly be affected, but no quantitative estimates  
for such stars
are yet available.  In what follows we will adopt the \citet{meynet} models. 
The models assume single star evolution, but even if the GRB progenitor were part of
a binary system, this would have no major impact on the lifetime which is
dominated by the main sequence phase.


\subsection{Spectral synthesis} 
Integrated photometry  is commonly analysed using spectral evolutionary synthesis models. 
In this procedure, it is important to adopt a model as closely adapted to the known physical properties of the target system as possible. As the GRB 030329 host galaxy is a blue object with strong emission-lines \citep{gorosabel05}, nebular emission is likely to give an important contribution to the optical spectrum.
Many publicly available models do not include nebular emission and are therefore not suitable for analysing systems of this kind. The previous GRB 030329 host galaxy analysis by \citet{gorosabel05} is  based on the \citet{Bruzual & Charlot} model, which suffers from 
this shortcoming.

In the following, we adopt the \citet{Zackrisson et al. a} spectral evolutionary model, which uses the photoionisation code Cloudy version 90.05 \citep{Ferland et al.} to predict the nebular continuum and emission lines.  For each time step, the spectral energy distribution of the  stellar population is used as input to Cloudy, which makes the computation time-consuming, but provides a realistic  evolution of the nebular component. The stellar component, which includes pre-main sequence evolution and a stochastic treatment of horizontal branch morphologies at low metallicities, is based on synthetic stellar atmospheres by  \citet{Lejeune et al.} and \citet{Clegg & Middlemass}, together with stellar evolutionary tracks mainly from the Geneva group. 
We have previously demonstrated \citep{ostlin2003} that the \citet{Zackrisson et al. a} model performs much better than the publicly available version of the P\'EGASE.2 code \citep{Fioc & Rocca-Volmerange} in reproducing the observed colours of young super star clusters. 
(Although P\'EGASE.2 has previously been used in the analysis of long-duration GRB host, e.g.
\citealt{Sokolov et al.,sollerman05}.) This is thanks to its more sophisticated treatment of the nebular component which is known to contribute substantially. 
To allow an analysis of the broadband magnitudes relevant for the current investigation, the spectra predicted by the models are redshifted to $z=0.1685$ and convolved with the throughput curves of the filters used.

In the \citet{Zackrisson et al. a} model, the properties of the stellar component are regulated by the metallicity, initial mass function (IMF) and star formation history. We here assume the metallicity  to be the same for the stars and the ionized gas, i.e. $Z=Z_\mathrm{stars}=Z_\mathrm{gas}$ and have explored $Z=0.001, 0.004$ and 0.008. For IMF we have assumed a power-law ($\mathrm{d}N/\mathrm{d}M\propto M^{-\alpha}$) throughout the 0.08--120 $M_\odot$ stellar mass range, and we note that the actual slope at low masses ($M <1$\msun ) will not affect the results for the young ages here considered. Our standard assumption is a Salpeter ($\alpha =2.35$) IMF but we have also explored $\alpha=1.85$ and 2.85. For simplicity, we assume the star formation rate (SFR) to be either instantaneous  or exponentially decreasing over time ($\mathrm{SFR}(t)\propto \exp{-t/\tau}$), where the e-folding decay rate $\tau$ describes the duration of the star formation episode. An instantaneous burst corresponds to $\tau \approx 0$.

The properties of the nebular component are determined by the adopted hydrogen number density $n(\mathrm{H})$, which is assumed to be constant within the nebula, the gas filling factor $f$,  and the gas mass $M_\mathrm{gas}$ available for star formation throughout the star formation episode. 
$M_\mathrm{gas}$ is used together with the adopted SFR($t$) to determine the stellar Lyman continuum flux density for each time step. This becomes an important parameter, because the homology relation between plane-parallel nebulae with the same ionisation parameter \citep{Davidson} breaks down for the spherical nebulae assumed here.

The \citet{Zackrisson et al. b} model assumes that the nebula responds instantaneously to changes in the ionizing flux from the stellar population. This approximation holds as long as the many processes in the ionized cloud occur on timescales shorter than the age resolution that we are interested in. 
To ensure that this is not an issue, we  used Cloudy to verify that the timescales of the longest radiative 
processes in the ionised gas are shorter than the time resolution of the stellar population model. For the 
range of gas parameters used in this paper, all processes occur on timescales much lower than 1 Myr, which 
is quite sufficient for our purposes.

\begin{figure*}
\includegraphics[width=84mm]{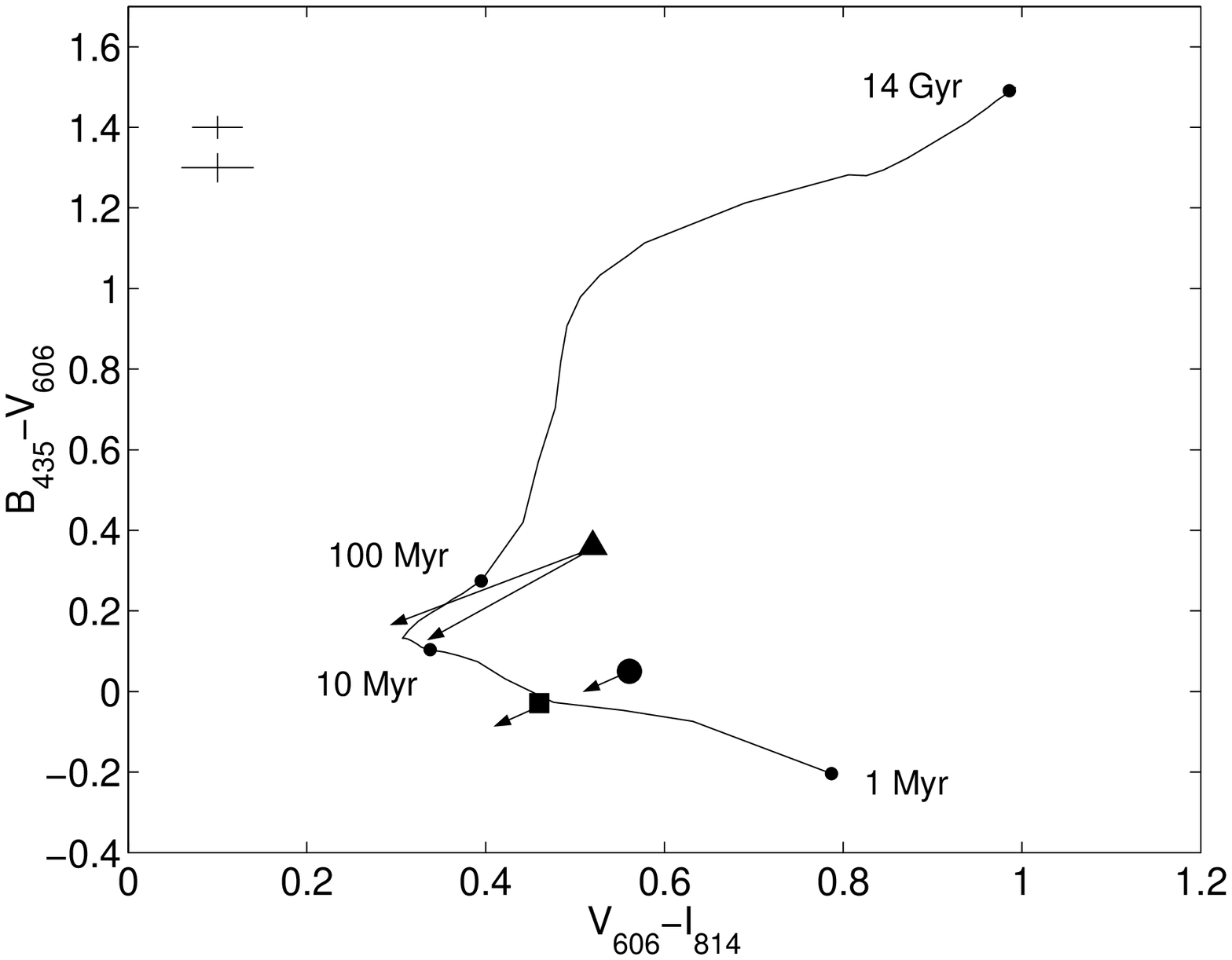}
\includegraphics[width=84mm]{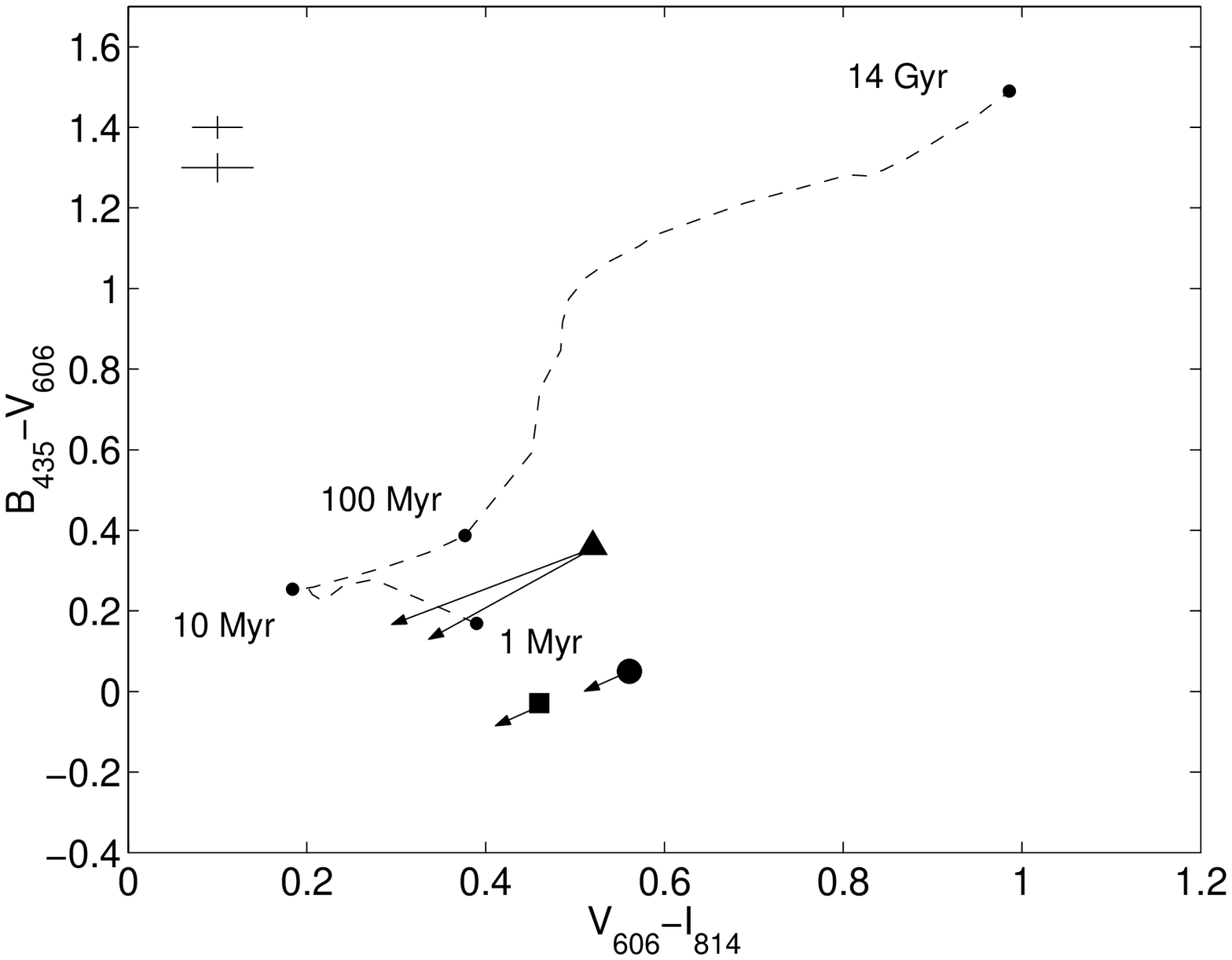}
\caption{The $V_{606}-I_{814}$ vs. $B_{435}-V_{606}$ colours of the integrated GRB\,030329 
host galaxy (filled triangle) and the GRB explosion site (filled square and circle,
indicating the colours with and without subtraction of the local background).
The arrows indicate the dereddening vectors for Milky Way and Small Magellanic
Cloud extinction curves \citep{Pei}, given the extinction estimates of
\citet{gorosabel05} for the overall galaxy and \citet{matheson03} for
the explosion site. At this resolution, the two arrows overlap in the latter
case. The smaller cross in the upper left corner indicates the 1\,$\sigma$
statistical uncertainties of the host galaxy  colours, whereas the larger cross indicates the  uncertainties
for the GRB explosion site. {\bf Left:} The solid line represents
 the spectral evolution predicted for a burst-like stellar population at $z=0.1685$ with
$Z=0.004$, $\tau=10^8$ yr, $M_\mathrm{gas}=10^8 \ M_\odot$, $f=0.001$ and
$n(\mathrm{H})=10^2$ cm$^{-3}$. The dots along this line indicate ages
of 1 Myr, 10 Myr, 100 Myr and 14 Gyr. After
reddening correction, the host galaxy is consistent with an age of $\sim 50$
Myr. {\bf Right:} The dashed line represents the predicted evolution for a
population with $\tau=5\times 10^7$ yr and $n(\mathrm{H})=10$ cm$^{-3}$, but
parameter values otherwise identical to those in the left hand panel. In
this case, the best-fitting age becomes only 1--2 Myr, demonstrating the
degeneracies involved in determining accurate ages for young stellar populations
from optical broadband photometry only.}
\label{BVI_galfig}
\end{figure*}

\subsection{The GRB 030329 host galaxy}
 The host galaxy of GRB\,030329  has been investigated in several studies using ground based data. 
\citet{gorosabel05} used spectroscopy of the host to estimate a metallicity of $Z=0.004$. However, 
\citep{sollerman05} found that the available spectroscopy in the literature does not strongly constrain the 
metallicity of the host galaxy, and neither $Z=0.001$ nor $Z=0.008$ could be ruled out. \citet{gorosabel05} could 
not tightly constrain the extinction but found that the host galaxy spectral energy distribution and Balmer line 
ratios favoured  $A_V \le 0.6$. \citet{matheson03} used the colour of the GRB afterglow to infer $A_V=0.12\pm 0.22$. 
Extinction may vary spatially over  a galaxy, and the GRB environment may well have a substantially lower 
extinction than that of the luminosity-weighted average of the host galaxy. This scenario also turns out to be consistent with 
our modelling results presented in Sect. 4.3.

In Fig.~\ref{BVI_galfig}, we compare the colours of the overall host galaxy (filled triangle) to those of the GRB explosion site (filled square and circle, indicating colours with and without subtraction of the local background, respectively). 
While similar in $V_{606}-I_{814}$, the $B_{435}-V_{606}$ colour measured at the explosion site is bluer than for the host galaxy   by $\sim 0.3$ magnitudes.
The dereddening vectors relevant for the overall host galaxy \citep{gorosabel05} and the GRB explosion sites \citep{matheson03} are plotted as arrows. For the host galaxy, the two arrows included correspond to the extinction curves for the Milky Way and Small Magellanic Cloud \citep{Pei}, respectively. For the GRB site, only one arrow has been plotted, because the two corrections overlap at this resolution. After correction, the GRB site is bluer than the host galaxy  in $B_{435}-V_{606}$ but redder in $V_{606}-I_{814}$. As will be demonstrated, this is consistent with the notion that the stellar population hosting the GRB progenitor is substantially younger than that of the overall host galaxy.

The solid line in the left panel of Fig.~\ref{BVI_galfig} indicates the evolution predicted for a $Z=0.004$,
$\tau=10^8$ yr, $M_\mathrm{gas}=10^8 \ M_\odot$ stellar population with gas parameters $f=0.001$ and $n(\mathrm{H})=10^2$ cm$^{-3}$. After reddening corrections, the host galaxy $B_{435}$, $V_{606}$ and $I_{814}$ data are consistent with a luminosity-weighted age of $\approx 50$ Myr. This corresponds to the lifetime of a $\sim8 \ M_\odot$ star, which is close to the minimum progenitor mass for a core-collapse SN \citep{2003ApJ...591..288H,2004MNRAS.353...87E}. 
The collapsar scenario for long-duration GRBs requires a higher progenitor mass than this; a minimum mass of $\sim$25--30 $M_\odot$ is required in order for stars to have cores that collapse to black holes and to have sufficient angular momentum to form a disk  \citep{2003ApJ...591..288H}.

Age estimates for composite stellar populations based on optical broadband photometry alone are typically not very precise. This is due to the degeneracy between star formation history and age \citep[e.g.][]{Gil de Paz & Madore,Zackrisson et al. b}, and, for systems with ongoing star formation, because of uncertainties in the physical conditions of the ionised interstellar medium \citep[e.g.][]{Zackrisson et al. a}. These effects are demonstrated in the right-hand panel of Fig.~\ref{BVI_galfig}, where the $B_{435}$, $V_{606}$ and $I_{814}$ data are compared to the predicted evolution for a model identical to the one in the left panel, except for a slightly shorter star formation episode ($\tau=5\times 10^7$ yr) and a lower hydrogen  density ($n(\mathrm{H})=10$ cm$^{-3}$). In this case, the best-fitting age becomes as small as 1--2 Myr. 
The large difference between the evolution in the two panels of Fig. 2 at low ages is entirely due to changes 
in the nebular spectrum. Since we have little information on the exact star formation history of the host galaxy and its detailed gas properties, the optical broadband data simply do not contain sufficient 
information to constrain the age with very high precision. 

\begin{figure}
\includegraphics[width=84mm]{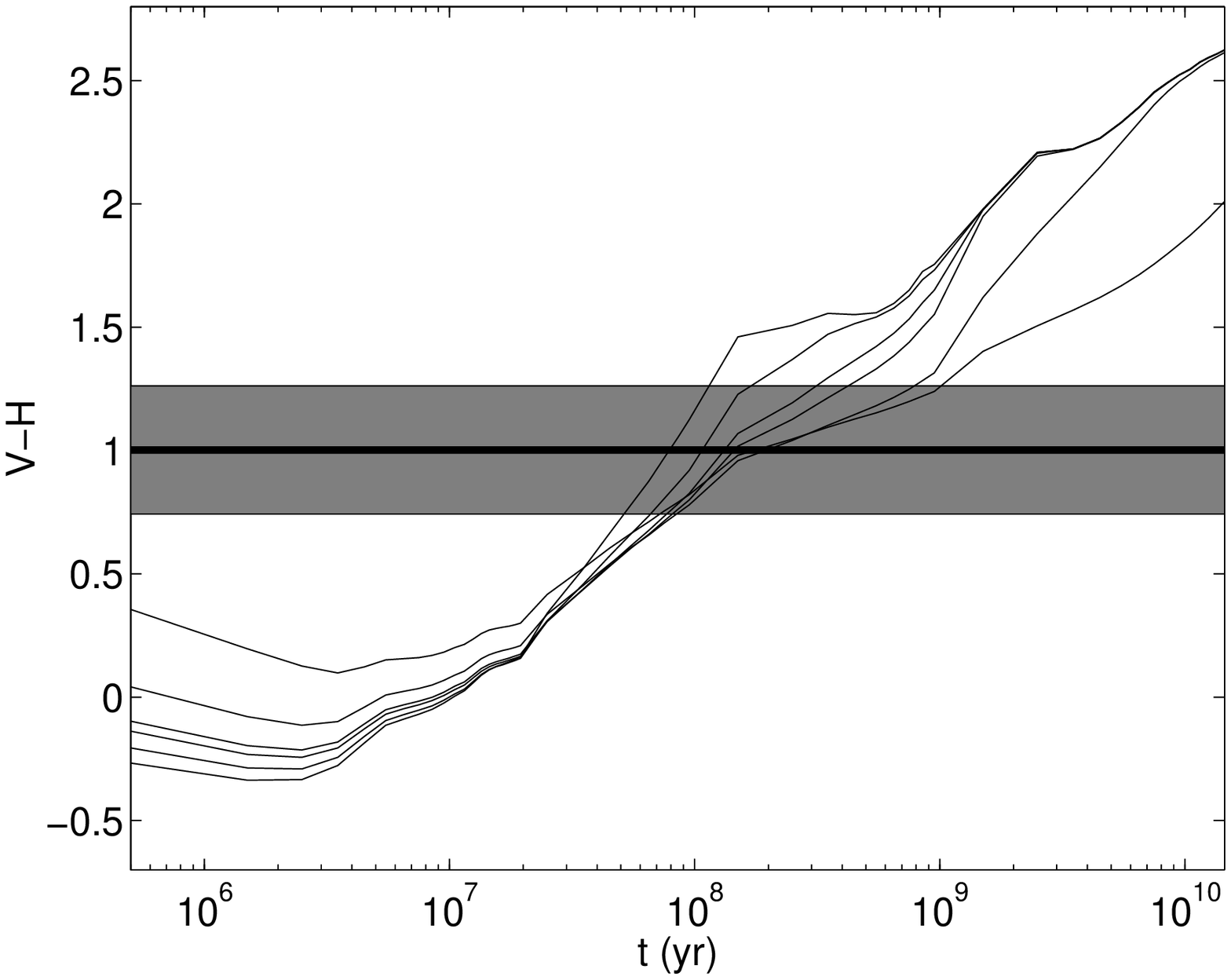}
\caption{The temporal evolution in $V-H$ predicted by various models compared to the ground-based data by \citet{gorosabel05} for the host galaxy (thick solid line, with grey region indicating 1\,$\sigma$ error bars). The observed colour has been corrected for Galactic and intrinsic extinction -- the latter assuming the \citet{gorosabel05} extinction estimates and a Small Magellanic Cloud extinction curve. 
The different model predictions correspond to star formation histories $\tau=5\times 10^7$, $10^8$, $2\times 10^8$, $3\times 10^8$, $10^9$, $10^{10}$ yr (from top to bottom at an age of $\sim 5\times 10^8$ yr). A population with $Z=0.004$, $M_\mathrm{gas}=10^8 \ M_\odot$, $f=0.01$ and $n(\mathrm{H})=10^2$, cm$^{-3}$ has been assumed throughout. Regardless of model scenario, the best-fitting age lies around $\sim 100$ Myr.}
\label{VH_galfig}
\end{figure}

An age higher than 50 Myr is nonetheless favoured by the ground based near-IR data of \citet{gorosabel05}. 
This is illustrated in Fig.~\ref{VH_galfig}, where the temporal evolution of various model sequences are 
compared to their $V-H$ measurement for the GRB\,030329 host galaxy.
In all cases, the best-fitting ages lie around $\approx 100$ Myr, which is broadly consistent with \citet{gorosabel05}. 
The ages inferred from $V-H$ are relatively robust to changes in the assumed nebular gas parameters.

While a best-fitting age of $\sim 100$ Myr would correspond to a GRB progenitor mass much lower than that 
favoured by the collapsar scenario, stars with much younger ages -- and hence higher masses -- would still be
present in a population with on going star formation (i.e. $\tau > 0$). Hence, while the integrated host galaxy properties are broadly consistent with a collapsar, they do not provide useful constraints on the mass of the actual GRB progenitor. 

\subsection{The GRB explosion site and the age of the GRB progenitor}\label{discussion}
\begin{figure*}
\includegraphics[width=84mm]{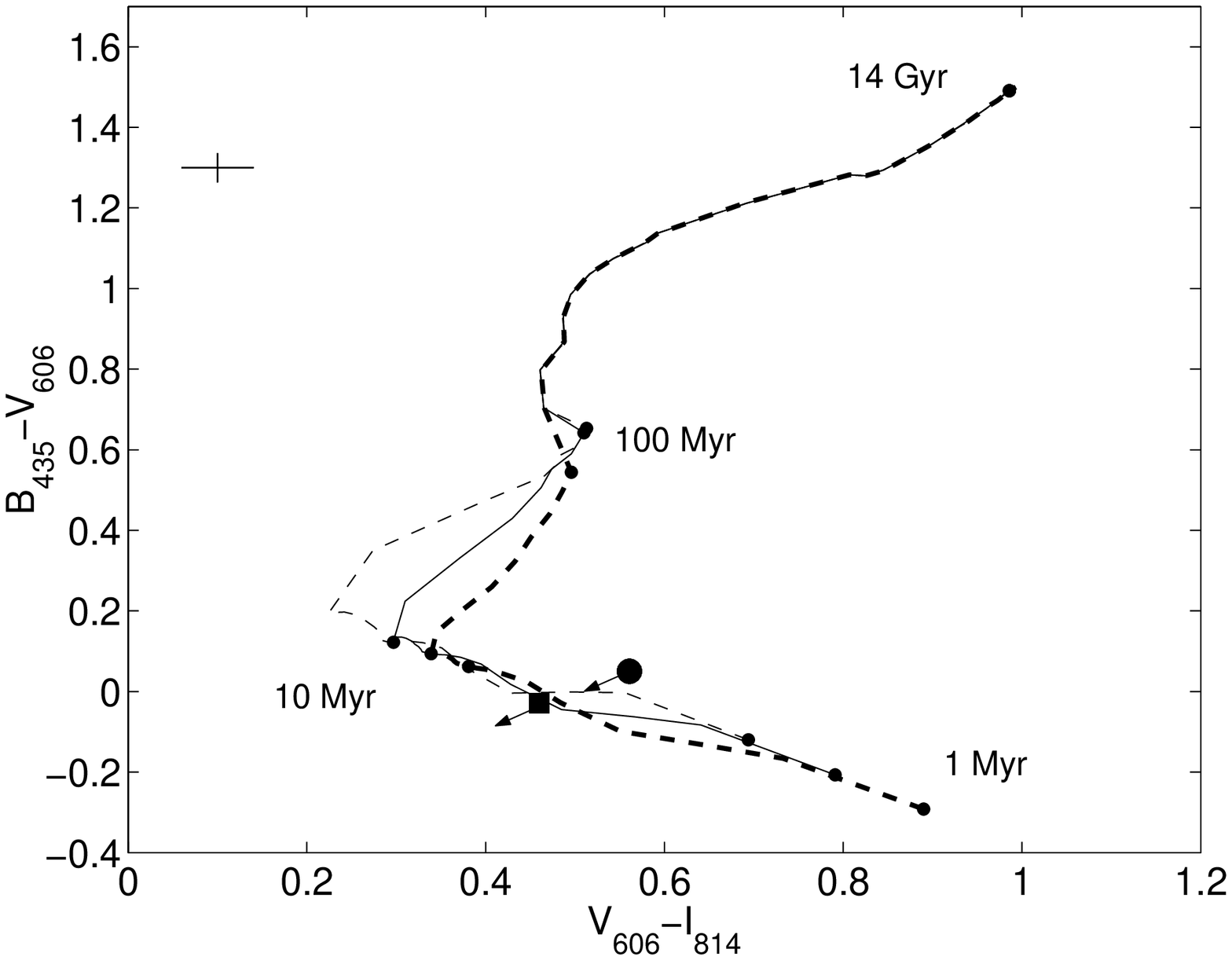}
\includegraphics[width=84mm]{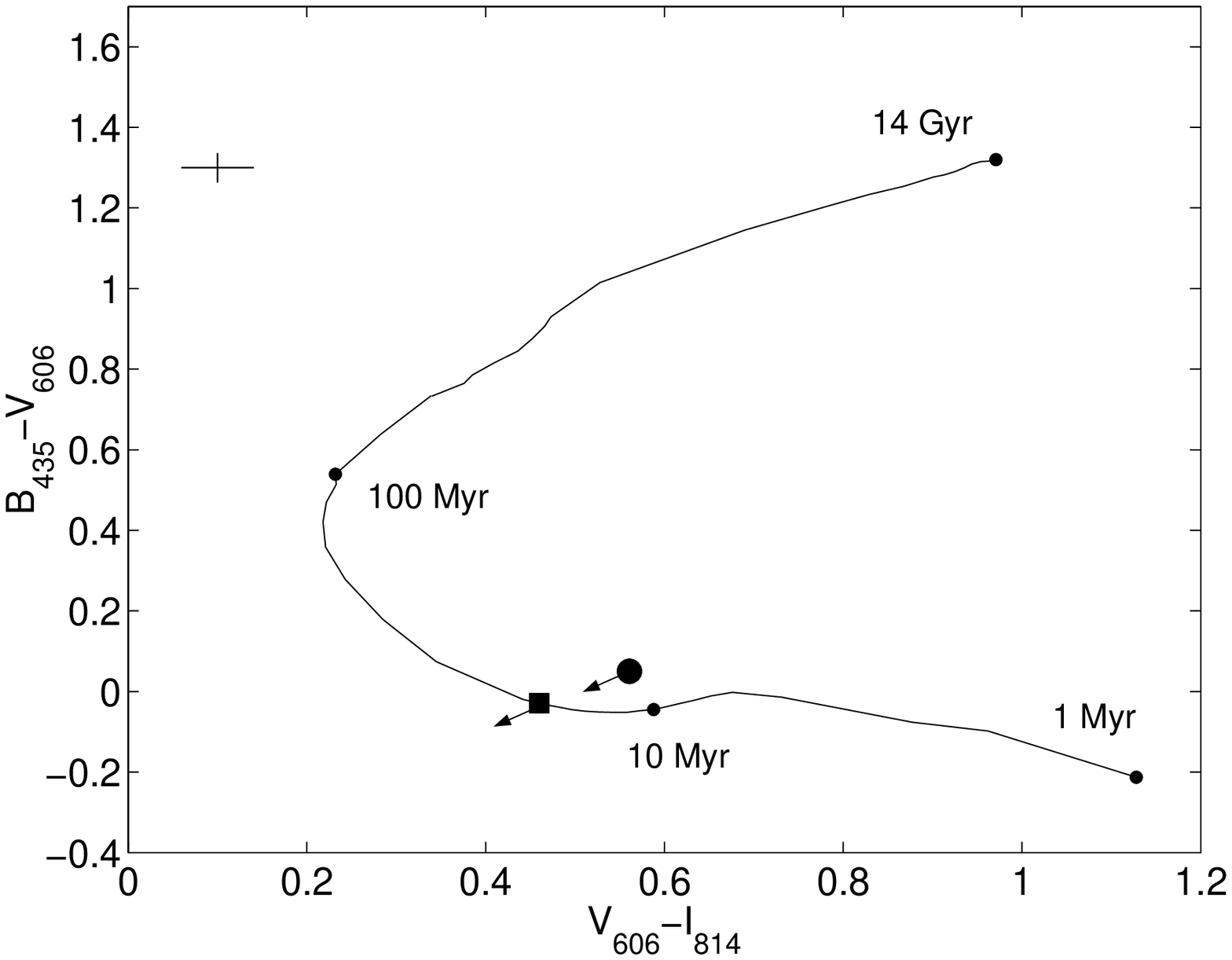}
\caption{The $V_{606}-I_{814}$ vs. $B_{435}-V_{606}$ colours of the GRB 030329 explosion environment 
(as in Fig. 2, filled square and circle represent colours derived with and without subtraction of 
the local background) compared to  various model populations (lines). 
 The extinction corrections are the same as in Fig.~\ref{BVI_galfig}.
{\bf Left:} Stellar populations with exponentially decreasing star formation rates of $\tau=5\times 10^6$ (thin dashed), $10^7$ (thin solid) and $2\times 10^7$ yr (thick dashed). All scenarios assume $Z=0.004$, $M_\mathrm{gas}=10^7 \ M_\odot$, $f=0.001$ and $n(\mathrm{H})=10^2$ cm$^{-3}$, leading to a best-fitting age of less than 10 Myr.
{\bf Right:} A stellar population with $Z=0.001$, but with parameters otherwise identical to the $\tau=10^7$ yr scenario from the left hand panel. In this case, the best-fitting age becomes $\approx 15$ Myr.}
\label{BVI_burstcol}
\end{figure*}

Starburst regions often contain  young star clusters, sometimes referred to as 
'super star clusters' (SSCs) with absolute magnitudes $M_V<-10$ \citep{arp,meurer}. At closer look, luminous starburst regions are
often composed of numerous SSCs of varying age \citep{ostlin2003}. Each such SSC must have 
formed on a short timescale \citep{mckee} and they are commonly assumed to be truly 
single stellar populations, even if the star formation history of a whole starburst region 
may be more complex. A nearby example is R136, the central cluster of the 30 Doradus complex
which \citet{mh} found to be well explained by a 
star formation episode of duration up to 4 Myrs and 
a standard Salpeter IMF. 

The  restframe absolute magnitude of the GRB explosion site is $M_{V_{606}}\approx-14.8$  
which
indicates a stellar population mass of $\sim 10^6\ M_\odot$ (the exact value depends on the actual
age and reddening). Our extraction radius (0.1\arcsec) corresponds to 0.28 kpc and the GRB exploded at the eastern edge of 
the highest surface brightness region in the starburst \citep[see Fig.~1 and also][]{sollerman2005}. 
Hence it may  have occurred in a single very luminous SSC, or in a complex of several SSCs of more 'normal' luminosity.

If the GRB occurred in a single SSC, then the star formation time scale ($\tau$) should be
close to instantaneous (or have a maximum $\tau$ of a few Myr), after which negative feedback associated 
with photo-ionisation and stellar winds would have quenched
 further star formation.
However, our aperture may include more
than one SSC in projection, and their different ages could then mimic a temporally extended star 
formation event. In the nearby blue compact galaxy ESO\,338--04, the central 2 kpc
starburst is composed of more than 50 young SSCs with ages ranging
from new-born to  about 40 Myr \citep{ostlin2003}. If the situation is similar in the host 
galaxy of GRB\,030329, then the effective star-formation time-scale $(\tau)$ in our 
aperture could be extended up to 10--20 Myr. This uncertainty on $\tau$
is the most serious limitation for our method and is directly tied to the spatial
resolution and distance of the target. Below we shall discuss the constraints on
the age of the GRB progenitor for various values of $\tau$ and other parameters.

The left panel of Fig.~\ref{BVI_burstcol} shows that the GRB 
explosion site can be well-fitted by a $Z=0.004$, $M_\mathrm{gas}=10^7 \ M_\odot$ 
population with a star formation time scale around $\tau=10^7$ yr. 
Contrary to the case for the overall host galaxy, the age of the GRB environment 
appears to be very young -- well below 10 Myr 
(corresponding to $M_\mathrm{ZAMS}> 21\ M_\odot$). 
Just as in the case of the host galaxy, the exact age is uncertain because of 
the aforementioned effects related to star formation history and properties of the 
ionized interstellar medium. In Fig.~\ref{BVI_burstcol}, three different star formation 
histories are plotted ($\tau=5\times 10^6$, $10^7$ and $2\times 10^7$ yr).  All of 
these scenarios provide reasonable fits to the observed colours of the GRB site, 
but at slightly different ages (2--5 Myr). 

The variations in the colours predicted by these scenarios at such early stages of 
evolution come from the different Lyman 
continuum flux densities that these stellar populations produce. 
While the 
best-fitting ages  for the GRB environment are systematically lower than those 
for the overall host galaxy, it is still possible to find a small number 
of model parameter combinations  that would allow  higher than 10 Myr. This is 
demonstrated in the right panel of Fig.~\ref{BVI_burstcol}, where the metallicity 
of the $\tau=10^7$ yr scenario from the left panel has been changed to $Z=0.001$, 
giving an age of $14$ Myr. This is actually the highest best-fitting 
age found when fitting the $B_{435}$, $V_{606}$, $I_{814}$ and $J_{110}$ magnitudes 
of the GRB environment to the 100  
sets of model parameter values explored. It corresponds to 
a GRB progenitor mass of $M_\mathrm{ZAMS} \geq 15 \ M_\odot$. 
We once again stress, that because of the adopted $\tau>0$ star formation history, 
the GRB progenitor may in this case have a lifetime anywhere in the range from 3 to 
14 Myr, i.e. fully consistent with the collapsar scenario of \citet{2003ApJ...591..288H}. 

\begin{figure}
\includegraphics[width=84mm]{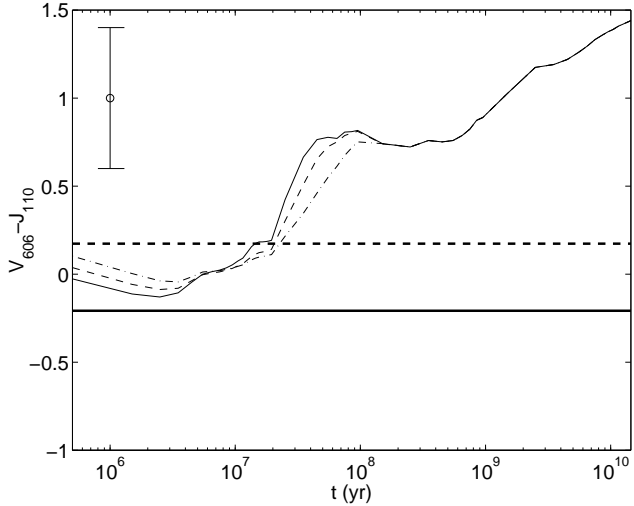}
\caption{The temporal evolution in $V_{606}-J_{110}$ predicted by various model scenarios compared to the colour of the GRB explosion site with (thick solid line) and without (thin dashed line) subtraction of the local background. The open symbol in the upper left indicates the statistical 1\,$\sigma$ uncertainty. The observed colours have been corrected for Galactic and intrinsic extinction -- the latter assuming the \citet{matheson03} estimate and a Small Magellanic Cloud extinction curve. 
The difference between Milky Way, Small and Large Magellanic Cloud extinction 
curves \citep{Pei} only amount to an uncertainty of 0.02 magnitudes in this colour. 
The different model predictions correspond to star formation histories $\tau=5\times 10^6$ (solid line), $10^7$ (dashed) and $2\times 10^7$yr (dash-dotted). A population with $Z=0.004$, $M_\mathrm{gas}=10^7 \ M_\odot$, $f=0.001$ and $n(\mathrm{H})=10^2$, cm$^{-3}$ has been assumed throughout. Due to the uncertainty in the subtraction of the local background, a best-fitting age of anywhere from 0 to 20 Myr can be derived, and within the error bars, ages of up to 100 Myr are in fact allowed. Such high ages are however inconsistent with the $B_{435}-V_{606}$ and $V_{606}-I_{814}$ colours.}
\label{VJ_burstcol}
\end{figure}

As shown in Sect. 4.2, near-IR data can be very useful for improving age estimates 
based solely on optical photometry. 
In Fig.~\ref{VJ_burstcol} we compare the $V_{606}-J_{110}$ colours of the GRB environment 
with the evolution predicted for different values of $\tau$. 
Colours with (thick solid line) and without (thick dashed line) subtraction of the local background are shown.
Due to the large photometric uncertainties, a best-fitting 
age of anywhere from 0 to 20 Myr can be derived from this colour.
The quality of the $J_{110}$ data is obviously too poor 
 to substantially improve the age estimates.  

In all, 100 models  were fitted to the data in Table 2. 
The variations in colours between the different photometry methods give a good
representation of the systematic uncertainties, whereas the statistical uncertainties
are always small, with the exception of the $J_{110}$-band. We compare the various
photometries (thus spanning the range of systematic uncertainties) to the models
and select those solutions that match the data to within 3\,$\sigma$ of the 
statistical uncertainties. 
The extinction has been a free parameter in the fit, but all fits within 3\,$\sigma$ 
of the photometric uncertainties consistently produced small values: $E(B_{435}-V_{606})<0.1$ mag.
This is consistent with the value derived from the optical afterglow \citep{matheson03},
which is reassuring.

For the 40  instantaneous burst models explored, we find that all models 
that provide good fits (within 3\,$\sigma$ of the photometric uncertainties)  
indicate a very young progenitor: the best fit is always 
less than 5 Myr (implying $M_\mathrm{ZAMS} \ge 47$ \msun)  and the 3\,$\sigma$ upper limit is 8 Myr 
($M_\mathrm{ZAMS} \ge 26$ \msun). 

Following the discussion above, we now also consider 60 additional models  with
more extended star formation episodes, $\tau \le 50$ Myr. The best fitting ages 
are in general still very young with the vast majority at $\le 5$ Myr. However 
a small number (1 to 3, depending on which photometry is used) of models give 
decent fits for  ages 7--14 Myr ($M_\mathrm{ZAMS} > 30$ to 
15 \msun ), with 3\,$\sigma$ upper limits of 20 Myr ($M_\mathrm{ZAMS} > 12$ \msun ). 
One of these models has already been presented in Fig. 4 (right panel). 

Of all the 100 models with $\tau \le 50$ Myr, 30--50\%  (depending on which
photometry is used)  provide best fit solutions that are within the 
3\,$\sigma$ statistical limits.
On average the photometry that provides the best fit to the models is the 'Best 
disk' without  subtraction of the local background.
The vast majority of models that produce good fits give very young ages
($\le 5$ Myr).  However, since we cannot rule 
out the physical parameters for those few that give higher ages, 
we are left with a rather loose mass constraint for the progenitor. 

We  note that if we 
would have adopted the rather common approach of simply assuming an instantaneous
burst with fixed metallicity and standard IMF and gas parameters, our constraints
would have appeared as very tight. Instead, we have conservatively explored a 
large parameter space in order to really test the power of this method.

\section{Outlook}\label{outlook}

As we have shown, the GRB\,030329 environment is best fitted by an age $\le 5$ Myr, 
which is  an order of magnitude lower than the age inferred for the overall host 
galaxy. This implies that progenitor mass estimates based on spectral or photometric 
data for poorly resolved host galaxies may be of limited value. Even with the 
high-resolution data analysed here, the uncertainties in the exact star formation history 
and gas properties of the GRB environment impose severe limits on the precision with 
which the mass of the GRB progenitor can be derived from broadband photometry. 

  An additional source of uncertainty stems from the intrinsic accuracy of
the spectral evolutionary model used. Currently, the only way to assess 
such uncertainties would be to analyze the data using several different 
models in parallel. As we have shown that the treatment of nebular emission is very 
important for the age derived for the GRB\,030329 environment, only models including 
a nebular component \citep[e.g.][]{Moy et al.,Anders & Fritze-v. Alvensleben,Magris et 
al.,Dopita et al.} should  be used in such a comparison.

To better constrain the GRB progenitor masses more diagnostics are required.
As already shown in Fig.~5, better quality near-IR data would be quite useful. 
This could be obtained with NICMOS, the next generation HST optical/IR imager 
WFC3, or with ground based adaptive optics systems.
It would also be very useful to complement the optical HST data with a $V$-like 
filter that does not transmit the [O{\sc iii}]$_{\lambda\lambda4959,5007}$ emission line, since this would 
make the analysis less sensitive to assumed nebular gas parameters.

\begin{figure}
\includegraphics[width=84mm]{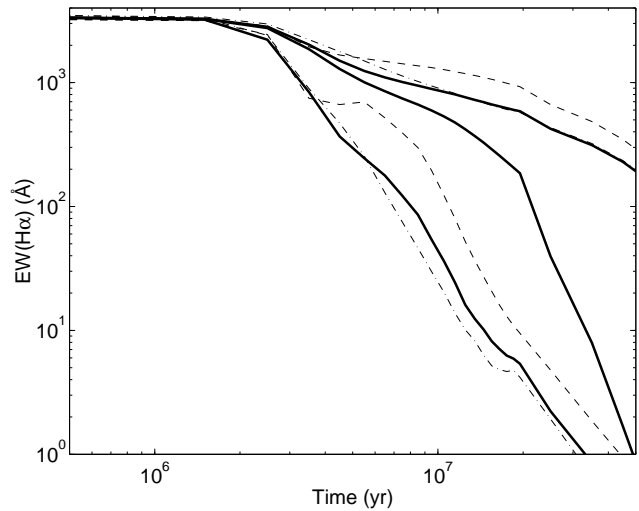}
\caption{The temporal evolution of the H$\alpha$ emission-line equivalent width
predicted for stellar populations with burst-like star formation histories. The
thick solid lines correspond to $Z=0.004$ populations with exponentially decaying
star formation rates of $\tau\approx 0$ (i.e. instantaneous burst), $5\times 10^6$
and $2\times 10^7$ yr (from bottom to top at an age of $10^7$ yr). 
The two thin dashed (dash-dotted) lines represent the corresponding equivalent width
predictions for $\tau\approx 0$ and  $\tau= 2 \times 10^7$ yr when instead $Z=0.001$
($Z=0.008$).}
\label{EWHa}
\end{figure}

The equivalent width of the H$\alpha$ emission line,  EW(H$\alpha$), measured  
through spectroscopy or narrow-band photometry, may also provide useful constraints. 
In Fig.~\ref{EWHa}, we present the EW(H$\alpha$) evolution predicted by the 
\citet{Zackrisson et al. a} model during the first 50 Myr  of a burst-like 
stellar population for three different star formation histories. For a given
 EW(H$\alpha$) measurement, the youngest age is inferred by the shortest star 
formation episode (here represented by an instantaneous burst), whereas the oldest 
age is inferred from the longest period of star formation  considered
($\tau = 2\times 10^7$ yr). Even without  detailed knowledge of the 
star formation history, interesting age  constraints can be derived from  EW(H$\alpha$). 
If, for instance, EW(H$\alpha$)$\approx 1000$~\AA, the allowed age range would be 3--8 Myr, 
corresponding to a progenitor star in mass the range 25--120 $M_\odot$. For EW(H$\alpha$)$\approx 100$ 
\AA, a lower limit of 8 Myr could be imposed, indicating a progenitor mass below 25 $M_\odot$. 
Moreover, the EW(H$\alpha$) of a young stellar population is almost completely 
insensitive to many of the parameters used in the \citet{Zackrisson et al. a} 
model to regulate the properties of the ionized gas.

While being a powerful age indicator, the H$\alpha$ equivalent width comes with a 
number of potential caveats that require consideration: While insensitive to the 
{\it gaseous} metallicity, there is a certain dependence on the  {\it stellar} 
metallicity. This is demonstrated in  Fig.~\ref{EWHa}, where we contrast the 
EW(H$\alpha$) evolution at $Z_\mathrm{stars}=Z_\mathrm{gas}=0.001$, 0.004 and 0.008. 
Without more detailed metallicity information than this, the 
age constraints derived from the EW(H$\alpha$)$\approx1000$ \AA\ and 100 \AA\ measurements 
would be weakened to 3--20 Myr and $\ge 7 $ Myr, respectively. 
Emission-line diagnostics formally probe the gaseous metallicity, which may differ 
somewhat from the stellar counterpart, and this will be a lingering source of error 
in the conversion from EW(H$\alpha$) to population age.

Selective extinction, i.e. the possibility that the emission line and the 
continuum originate from different spatial regions with different dust content 
\citep[e.g.][]{Calzetti et al.}, represents another potential complication.

This investigation provides support for the origin of long duration GRBs in very massive stars. 
In \citet{sollerman05} we found a similar conclusion for  SN\,1998bw (GRB\,980425). 
The parent population of core collapse supernovae is expected to be biased towards ages
corresponding to the lifetime of stars with mass close to the lower limit for SNe ($\sim 8 M_\odot$), 
since for a normal IMF these are more common than very massive stars.  
Already the finding that two of the still few known SNe coincident with long duration GRBs 
seem to have higher masses suggests that the parent populations may be different. This is consistent 
with studies of the spatial distribution of GRBs and SNe within galaxies \citep{fruchter06}.
 It also agrees with the modeling of the SNe associated with GRBs, which typically indicate 
very massive progenitors \citep[e.g.][Table 1]{mazzali07}.
Admittedly more examples would be needed  since for individual cases there is
always a risk that  photometry is affected by limited spatial resolution and chance 
projections  such that the progenitor had a different age than its surroundings.
By  imaging the most nearby  GRB host galaxies with HST-like resolution, population studies 
of the explosion environment can provide statistical constraints on GRB progenitor masses. 
The method could also be used for supernovae that are too distant for single stars in 
their host galaxies to be resolved ($D>20$ Mpc) and has been used for constraining the progenitors of
the type IIP SN 2004dj \citep{2004ApJ...615L.113M,2005ApJ...626L..89W}  and the type Ic SN 2007gr 
\citep{crockett08}

\section{Summary}\label{conclusion}

We have used archival HST images to derive photometry for the explosion site of GRB\,030329. 
This  has been used together with spectral evolutionary synthesis models, including 
nebular emission and a wide variety of parameter settings,  to constrain the age of 
the stellar population. Taking this age to equal the lifetime of the GRB progenitor star,
we can use the tight relationship between initial mass and lifetime to weigh the progenitor.


Studying the photometry at the  resolution allowed by HST, the explosion site has  
an inferred age of $\le 5$ Myr for an instantaneous burst assumption. The 3\,$\sigma$ upper
limit of 8 Myr corresponds to a minimum progenitor 
mass of 25 \msun, in line with the predictions of the collapsar model for the origin of long 
duration GRBs \citep{2003ApJ...591..288H}.
 
The main limitation is related to  spatial resolution (for $z=0.1685$ one HST/ACS/WFC pixel 
corresponds to 140 pc) and arises because we cannot be sure that the GRB 
parent population can be described by a single stellar population. Thus, nearby clusters of 
different age may contaminate the photometry, thereby mimicking an extended star formation history. 
Allowing for a more extended star formation history, most models still predict a minimum 
progenitor mass of around 25 \msun , but for a small number of models the minimum allowed 
(within 3\,$\sigma$) progenitor mass is  $\le 12$\msun.

We discuss ways to obtain tighter age constraints 
and identify the equivalenth width of H$\alpha$ and near-IR photometry as useful additional diagnostics. 

If exercised on a sufficient number of long duration GRBs, the method in this paper could provide information on the progenitor mass and metallicity distributions. This would provide an independent comparison to estimates obtained through modeling of the associated supernova.

\section*{Acknowledgements}
G.\"O. acknowledges support from the Swedish Research Council (VR) and the Swedish National Space Board.
E.Z. acknowledges  support from VR, the Royal Swedish Academy 
of Sciences (KVA) and the visitor programme at the Dark Cosmology Centre, 
where a large part of this work was carried out. 
The Dark Cosmology Centre is funded by the Danish National Reserach Foundation.
G.\"O. and J.S. acknowledges a grant from the Crafoord
foundation, administrated through KVA.
The work of S.M., conducted as part of the award "Understanding the lives of massive stars from
birth to supernovae"  made under the European Heads of Research Councils and European Science
Foundation EURYI (European Young Investigator) Awards scheme, was supported by funds from the
Participating Organisations of EURYI and the EC Sixth Framework Programme, and also the Academy of
Finland (project: 8120503).

\label{lastpage}
\end{document}